\documentclass[prb,aps, babel,twocolumn]{revtex4}
\usepackage{amsmath,amssymb,amstext,amscd,amsthm,amsopn,graphicx,indentfirst,mathrsfs}

\newcommand{\w}{\omega}

\graphicspath{{.}{./EPS/}}

\begin{document}

\title{Resistivity of Inhomogeneous Superconducting Wires}

\author{G. Venketeswara Pai$^{1,2}$, E. Shimshoni$^2$ and N. Andrei$^3$}

\affiliation{
$^1$Department of Physics, Technion, Haifa 32000, Israel\\
$^2$Department of Mathematics--Physics,
University of Haifa at Oranim, Tivon 36006, Israel\\
$^3$Center for Materials Theory, Rutgers University, Piscataway, NJ
08854, USA}

\date{\today}

\begin{abstract}
We study the contribution of quantum phase fluctuations in the
superconducting order parameter to the low--temperature resistivity
$\rho(T)$ of a dirty and inhomogeneous superconducting wire. In
particular, we account for random spatial fluctuations of arbitrary
size in the wire thickness. For a typical wire thickness above the
critical value for superconductor--insulator transition,
phase--slips processes can be treated perturbatively. We use a
memory formalism approach, which underlines the role played by weak
violation of conservation laws in the mechanism for generating
finite resistivity. Our calculations yield an expression for
$\rho(T)$ which exhibits a smooth crossover from a homogeneous to a
``granular'' limit upon increase of $T$, controlled by a
``granularity parameter'' $D$ characterizing the size of thickness
fluctuations. For extremely small $D$, we recover the power--law
dependence $\rho(T)\sim T^\alpha$ obtained by unbinding of quantum
phase--slips. However in the strongly inhomogeneous limit, the
exponent $\alpha$ is modified and the prefactor is {\em
exponentially enhanced}. We examine the dependence of the exponent
$\alpha$ on an external magnetic field applied parallel to the wire.
Finally, we show that the power--law dependence at low $T$ is
consistent with a series of experimental data obtained in a variety
of long and narrow samples. The values of $\alpha$ extracted from
the data, and the corresponding field dependence, are consistent
with known parameters of the corresponding samples.
\end{abstract}


\maketitle

\section{Introduction}
\label{sec:intro}

Transport in superconducting systems of reduced dimensions (thin
films and wires) is known to be strongly affected by fluctuations in
the order parameter. A prominent manifestation of the role of
fluctuations is the finite electrical resistance of narrow
superconducting (SC) wires at any finite temperature $T$ below the bulk
critical temperature $T_c$, established when the wire thickness $d$
is reduced below the superconducting coherence length \cite{Little}
$\xi$. The finite voltage drop along the wires is generated by
phase--slips: these are processes whereby the superconducting phase
slips by $2\pi$ at points on the wire where superconductivity is
temporarily destroyed. The resulting voltage is related to the rate
of phase slips via the Josephson relation. The pioneering
theoretical studies of this phenomenon \cite{LAMH} accounted for
thermally activated phase--slips across barriers separating
metastable phase configuration states, corresponding to local minima
of the Landau--Ginzburg free energy. This theory turns out to be
supported by experimental studies \cite{TAPS}, in particular
providing a good fit of the resistivity $\rho(T)$ slightly below
$T_c$. As $T$ is lowered further, $\rho(T)$ is exponentially
suppressed and becomes practically undetectable.

The advance of nanostructure fabrication techniques during the
1980's opened up the possibility to study narrow wires of smaller
diameter, down to a few 100 {\AA}. The consequent weakening of
superconductivity leads to enhancement of the rate of phase--slips,
and results in a measurable finite resistance even at $T\ll T_c$. In
this low $T$ regime, however, thermal activation is considerably
suppressed, and the dominant mechanism for phase--slips becomes
quantum tunneling. The first experimental indication of quantum
phase--slips (QPS) has been seen in thin strips of Indium by
Giordano \cite{Giordano}. The curves of $\rho(T)$ vs. $T$ exhibit a
`kink' at some $T^\ast$ smaller than $T_c$, below which the decrease
of $\rho(T)$ upon lowering $T$ becomes more moderate. For
$T<T^\ast$, the fit to the theory of Ref.~[\onlinecite{LAMH}] fails.
Motivated by the assumption that quantum tunneling dominates over
thermal activation in this regime \cite{LegCal}, Giordano fitted the
data to a phenomenological expression in which the temperature $k_B
T$ in the thermal activation rate $\sim e^{-\Delta F/k_B T}$ is
replaced by a ``characteristic frequency'' given by
$\hbar/\tau_{LG}$, where $\tau_{LG}$ is the relaxation time in the
time--dependent Landau--Ginzburg theory. While this phenomenological
expression fitted the data quite successfully, it is not justified
by a rigorous theoretical derivation. In particular, the
Landau--Ginzburg theory (which is employed in the calculation of the
barrier hight $\Delta F$) is based on an expansion in the close vicinity of $T_c$,
and is not expected to hold far below $T_c$.

Subsequent theoretical studies of the QPS contribution to
resistivity below $T_c$
\cite{Duan,Renn,zaikin,zaikin_prb,khleb1,KP,RDOF} have addressed the
quantum dynamics of phase fluctuations in the low--temperature limit
($T\ll T_c$), yielding a radically different behavior of $\rho(T)$
vs. $T$. At low $T$, the magnitude of the SC order parameter field
is approximately constant and its dynamics is dominated by
phase--fluctuations. The corresponding $T\rightarrow 0$ effective
model is a (1+1)--dimensional XY--model, in which the topological
excitations are QPS and anti--QPS (vortices and anti--vortices in
space--time) \cite{zaikin}. Upon tuning a stiffness parameter (which
in particular is proportional to the wire diameter $d$) below a
critical value, the system undergoes a quantum Kosterlitz--Thouless
(KT)\cite{KT} transition from a SC phase to a metallic phase driven
by the unbinding of QPS--anti QPS pairs. The latter are responsible
for a finite resistivity at any finite $T$ even in the SC phase,
where
\begin{equation}\label{powerlaw}
\rho(T)\sim \rho_0 \left(\frac{T}{T_c}\right)^\alpha
\end{equation}
with $\alpha>0$. This power--law dependence was predicted both for a
granular wire composed of weakly coupled SC grains separated by a
tunnel barrier (a Josephson junction array) \cite{Renn,RDOF} and for
the case of a homogeneous wire in the dirty limit
\cite{zaikin,zaikin_prb,KP}; however, the exponent $\alpha$ and the
prefactor $\rho_0$ are different (the latter, in particular,
strongly depends on the microscopic details).

The quantitative estimate of $\rho_0$ and the consequent size of the
resistivity in realistic systems have been discussed in detail in
the theoretical literature, and led to a debate among the authors
\cite{DuanCom} regarding the relevance of the homogeneous wire limit
to the experimental system of Ref. ~[\onlinecite{Giordano}].
However, we are not aware of a direct attempt to test the validity
of the power--law scaling of $\rho(T)$ as an alternative to
Giordano's phenomenological formula. More recently, several
experimental groups \cite{tinkham1,tinkham2,tian,arut,altomare} have
reported QPS--induced resistivity in a variety of SC nanowire
samples with $d$ of order 100 {\AA} and below. The data in certain
samples also exhibit a transition to a normal (weakly insulating)
state below a critical wire diameter $d_c$, consistent with the
theoretical expectations. In the SC samples, the data was fitted by
an effective circuit which accounts for both quantum and thermally
activated phase--slips contributions to $\rho(T)$. Here as well,
Giordano's phenomenological expression was implemented to describe
the QPS term in the fitting formula, rather than a power--law $\sim
T^\alpha$ (see, however, the unpublished notes in Ref.
~[\onlinecite{altomare}] and [\onlinecite{arut2}]). The comparison
between theory and experiment is therefore not entirely settled.

In the present paper, we introduce a derivation of QPS--induced
resistivity which enables a systematic account of all possible
scenarios in a realistic experimental setup, and carry out a
comparison with the experimental data. We focus our attention on
dirty SC wires, and account for non--uniformity of arbitrary size in
the wire parameters, in particular the diameter $d$. Our approach
implements the memory function formalism, which directly relates the
mechanism responsible for generating finite resistivity to the
violation of conservation laws of the unperturbed Hamiltonian $H_0$
(see below), describing non--singular fluctuations. The formalism
also underlines the interplay of phase--slip processes and disorder.
We consider two distinct types of disorder: the first corresponds to
impurities in the underlying normal electronic state, characterized
by a mean free path $\ell\ll\xi$, and the second type is associated
with the random spatial fluctuations in the wire diameter on a
length scale of order $\xi$, whose size is characterized by a
``granularity parameter" $D$. We find that $\rho(T)$ exhibits a
smooth crossover from a homogeneous to a ``granular'' behavior upon
increase of $T$. The latter is expected to dominate in most of the
measurable range of $T$ due to an exponential enhancement of the
prefactor $\rho_0$ in Eq. (\ref{powerlaw}). In addition, we examine
the dependence of the exponent $\alpha$ on an external magnetic
field applied parallel to the wire. The model and the main steps of
our calculation are described in Sec. \ref{sec:model}; details of
our derivation of the resistivity within the memory approach are
given in Appendix A. In Sec. \ref{sec:exp} we show that the
power--law Eq. (\ref{powerlaw}) is consistent with the experimental
data obtained in a variety of samples, provided the wire is
sufficiently long. The values of $\alpha$ extracted from the data,
and the corresponding field dependence, are consistent with known
parameters of the corresponding samples. Our conclusions are
summarized in Sec. \ref{sec:conclude}.

\section{The Model and Derivation of Principal Results}
\label{sec:model}

We consider a long and narrow superconducting (SC) wire of length $L$ and
cross section $s=\pi (d/2)^2$, such that $L\gg \xi$ and $d\ll \xi,\lambda_L$
where $\xi$ is the superconducting coherence length and $\lambda_L$ the
London penetration depth. Fluctuations in the SC order parameter are therefore effectively
one--dimensional. As a first stage we assume that the wire is homogeneous, and the SC
material is in the dirty limit where the mean free path in the underlying electronic
system obeys $\ell\ll \xi$. An effective model in terms of the SC order parameter field is
obtained as a result of integrating over the electron fields\cite{zaikin,zaikin_prb}.
At low temperatures $T\ll T_c$ (where $T_c$ is the bulk SC transition temperature),
we assume that fluctuations in the magnitude of
the order parameter are suppressed, while the quantum dynamics of phase
fluctuations $\phi(x)$ is described by the Sine-Gordon
Hamiltonian\cite{zaikin}
\begin{eqnarray}\label{HSG}
H=H_0+\sum_n H_n^{ps}\; ,
\end{eqnarray}
where
\begin{eqnarray}\label{H0}
H_0&=&\frac{1}{2}\int_{-L/2}^{L/2} dx \left[ \frac{(2e)^2}{C}\Pi^2 +
\frac{sn_s}{4m} (\partial_x \phi)^2 \right] \; ,\\
H_n^{ps}&=&\frac{-2 y^nv}{a^2}\int_{-L/2}^{L/2} dx
\cos(2n\theta)\label{Hps}\; .
\end{eqnarray}
Here and throughout the rest of the paper we use units where $\hbar=k_B=1$;
$\Pi(x)$ is the field conjugate to $\phi(x)$, satisfying
$[\phi(x),\Pi(x^\prime)]=i\delta(x^\prime-x)$, which physically represent
fluctuations in the number of Cooper pairs and is related to the field
$\theta$ via $\partial_x\theta=\pi\Pi$. $C$ is the effective
capacitance per unit length, $n_s$ is the
(three-dimensional) superfluid density, and $e$, $m$ are the electron charge
and mass, respectively. In $H_n^{ps}$, $y=\exp\{-S_{core}\}$ is the fugacity of quantum
phase--slip (QPS), where $S_{core}$ is the action associated with the
creation of a single QPS (of winding number 1) due to the suppression of the
SC order parameter in its core; the characteristic velocity $v$ is given by
\begin{eqnarray}\label{v-def}
v&=& \left(\frac{se^2n_s}{mC}\right)^{1/2}\; ,
\end{eqnarray}
$n$ is the winding number (``charge'') of a QPS, and $a$ is a short distance cutoff
$\sim\xi$. Note that the Hamiltonian Eq. (\ref{HSG}) is
the same effective model as in Ref.~[\onlinecite{zaikin}] expressed in a dual
representation\cite{CLbook}.

The Hamiltonian $H_0$ describes the non--singular phase fluctuations, which are
characterized by a free mode (the Mooij--Sch{\"o}n mode\cite{mooij}) propagating
at a velocity $v$. It can be recast in the familiar Luttinger form
\begin{eqnarray}\label{HLL}
H_0&=& v\int \frac{dx}{2 \pi}
\left( K (\partial_x \theta)^2+\frac{1}{K} (\partial_x \phi)^2 \right)\; ,
\end{eqnarray}
where $v$ is defined in Eq. (\ref{v-def}) and
\begin{eqnarray}\label{K-def}
K &=& \frac{4}{\pi}\left(\frac{me^2}{sn_sC}\right)^{1/2}\; .
\end{eqnarray}
Note that we also assume the wire to be sufficiently long so that
$L$ is large compared to $v/T$, in which case it can be practically
taken to be infinite. The nature of the $T=0$ fixed point of $H$
[Eq. (\ref{HSG})] depends crucially on the Luttinger parameter $K$.
In particular, as noted by Ref.~[\onlinecite{zaikin}], the system
undergoes a quantum Kosterlitz-Thouless (KT)\cite{KT} transition
from a SC phase at $K<K_c$ ($K_c\approx 1/2$) to a metallic phase at
$K>K_c$. For given material parameters, the transition can be tuned
by varying the wire cross section $s$ below a critical value $s_c$,
related to $K_c$ through Eq. (\ref{K-def}). We hereon focus on the
SC phase corresponding to $s>s_c$, in which all the terms $H_n^{ps}$
[Eq. (\ref{Hps})] are irrelevant and the $T=0$ fixed point
Hamiltonian is $H_0$. Indeed, it describes a true SC state, where
the resistivity $\rho(T)$ vanishes in the limit $T\rightarrow 0$.

A key feature of the SC state is that the charge current, dominated
by the superconducting component
\begin{equation}\label{Je:main}
J_e=\frac{esn_s}{m}\int dx \partial_x \phi =
\frac{2e}{\pi}\frac{v}{K}\int dx \partial_x \phi\; ,
\end{equation}
is an {\em almost conserved} quantity. Formally, this is manifested
by the vanishing of its commutator with the low--energy Hamiltonian
$H_0$, which implies that in the absence of the phase--slip terms
$H_n^{ps}$, the current cannot degrade ($\partial_t J_e=0$) and
hence the resistivity vanishes. The leading contribution to
$\rho(T)$ can therefore be obtained perturbatively in $H_n^{ps}$. As
we show in detail in Appendix A, the calculation of $\rho(T)$ can be
viewed as a particularly simple application of the more general {\it
memory matrix} approach\cite{forster,woelfle,giamarchi,rosch}, which
directly implements this insight. In essence, this approach
incorporates a recasting of the standard Kubo formula for the
conductivity matrix (a highly singular entity in the case of an
almost perfectly conducting system) in terms of an object named a
``memory matrix". The latter corresponds to a matrix of decay-rates
of the slowest modes in the system, and is perturbative in the
irrelevant terms in the Hamiltonian, in particular all processes
responsible for degrading the currents and hence generating a finite
resistivity. The separation, in this approach, of the slow modes
generated by the irrelevant operators around $H_0$ from the fast
modes allows a controlled approximation as the temperature is
lowered and provides a lower bound on the conductivity \cite{jung}.

Our derivation of the d.c. electric resistivity (see Appendix A for
details) yields the following expression:
\begin{eqnarray}\label{rho-ee}
\rho(T)= \frac{M_{ee}}{\chi_{ee}^2}\; ,
\end{eqnarray}
where $\chi_{ee}$ is the static susceptibility
\begin{eqnarray}\label{chi-ee}
{\chi}_{ee}= \frac{1}{T L} \langle J_e|J_e\rangle
\end{eqnarray}
and $M_{ee}$ (the memory function) can be expressed, to leading
order in the perturbations $H^{ps}_{n}$, in terms of correlators of
the `force' operators
\begin{eqnarray}\label{Fps-ee}
F_{ps,n}^e = i[J_e,H_n^{ps}]\; ,
\end{eqnarray}
which dictate the relaxation rate of the current $J_e$ via
\begin{eqnarray}\label{dtJe}
\partial_t J_e=\sum_n F_{ps,n}^e\; .
\end{eqnarray}
This yields
\begin{eqnarray}\label{MMee}
M_{ee}(T) &\approx&  \sum_{n} M^{ee}_{ps,n}\; , \\
M_{ps,n}^{ee}&\equiv& \lim_{\w\rightarrow 0}\frac{ \langle
F_{ps,n}^e;F_{ps,n}^e \rangle^0_\w- \langle F_{ps,n}^e;F_{ps,n}^e
\rangle^0_{\w=0}}{i \w}\nonumber
\end{eqnarray}
in which $\langle F^e;F^e\rangle^0_{\w}$ is the retarded correlation
function, the expectation value being evaluated with respect to
$H_{0}$.

To leading order in perturbation theory, we evaluate the expectation
value in Eq. (\ref{chi-ee}) as well with respect to the low energy
Hamiltonian $H_0$. This yields
\begin{eqnarray}\label{chi}
\chi_{ee}\approx \frac{8e^2v}{\pi K}\; .
\end{eqnarray}
Using Eqs. (\ref{Je:main}), (\ref{Hps}) and (\ref{Fps-ee}) we obtain
an expression for the force operator
\begin{eqnarray}\label{Fee}
F_{ps,n}^{e} = 8n \: \frac{e} {a^2}\: \frac{v^2} {K}\: y^n \int dx
\sin \left(2n \theta \right)\; .
\end{eqnarray}
Inserting in Eq. (\ref{MMee}), we find
\begin{eqnarray}\label{Mee}
M_{ps,n}^{ee} &=&  \frac {4L(4nev^2y^n)^2}{a^4K^2} \nonumber\\
&\times& \int_{-\infty}^{\infty} dx\int_{0}^{\infty} dt\, t\,{\rm
Im}\{C_{ps,n}^{ee}(x,t)\}\; ,
\end{eqnarray}
where
\begin{eqnarray}\label{Cee}
C_{ps,n}^{ee}(x,t)&=&e^{4n^2G_\theta(x,t)}\; , \\
G_\theta(x,t)&\equiv & \langle  \theta(x,t)\theta(0,0)\rangle^0 \nonumber
\end{eqnarray}
in which the Green's function $G_\theta(x,t)$ at finite $T$ is given
by \cite{schultz,Gbook}
\begin{eqnarray}\label{Green-T0}
 G_\theta(x,t)&=& \frac{1}{4K} \ln \left[\frac{\pi aT/v}
{\sinh\{\pi T(x-v t+ia)/v\}}\right] \\
\, &+& \frac{1}{4K}\ln \left[\frac{\pi aT/v}
{\sinh\{\pi T(x+v t-ia)/v\}}\right]\; . \nonumber
\end{eqnarray}
To find the leading $T$--dependence of $\rho(T)$ for small $y$ and $(aT/v)$,
we neglect the contributions of $n>1$. Substituting the resulting $M_{ee}$
combined with $\chi_{ee}$ from Eq. (\ref{chi}) in Eq. (\ref{rho-ee}), we obtain
\begin{eqnarray}\label{rho-hom-final}
\rho(T)&\approx &\frac{4\pi^3}{(2e)^2a}\frac{\Gamma^4(1/2K)}{\Gamma^2(1/K)}y^2
\left[\frac{2\pi aT}{v}\right]^{2/K-3}
\end{eqnarray}
where $\Gamma(x)$ is the Gamma function. This essentially recovers
(up to a numerical prefactor) the result of
Ref.~[\onlinecite{zaikin}], which indeed corresponds to the
homogeneous wire limit. Note that the resulting $\rho(T)$ exhibits a
SC behavior as long as $K<2/3$, in accord with the renormalization
group analysis of Ref.~[\onlinecite{GS}].

We next turn our attention to the more realistic situation, where inhomogeneities
along the SC wire are allowed. Random fluctuations are possible
in all the wire parameters, and in particular the diameter may varry in space
leading to a local cross section $s(x)$, which can be assumed to be a random function of $x$.
The most prominent modification of the Hamiltonian in the presence of such spatial fluctuations
is manifested in the fugacity $y$, which depends exponentially on the wire diameter via the core
action $S_{core}$. Consequently, it becomes space--dependent, i.e. $y=y(x)$ where
\begin{equation}\label{ydis}
y(x) = \exp \left(-S_{core} (x) \right) = \exp \left(-S_0 - \Lambda(x) \right)
\end{equation}
where $\Lambda(x)$ is a random correction to the uniform core action
$S_0$. The leading ($n=1$) phase--slip contribution to the
Hamiltonian [Eq. (\ref{Hps})] now becomes
\begin{eqnarray}\label{Hdis}
H_{ps}&=&\frac{-2 v}{a^2}\int dx \, y(x)\cos(2\theta) \nonumber
\\&=&\frac{-2 y_0v}{a^2}\int dx \, \exp\left(- \Lambda(x)
\right)\cos(2\theta)\; ,
\end{eqnarray}
where $y_0=\exp(-S_0)$. Assuming a Gaussian distribution of the
random function $\Lambda(x)$
\begin{equation}
P[\Lambda]={\cal N}\exp \left\{- \frac{1}{2aD}\int dx \Lambda^2(x)
\right\}
\end{equation}
yields the disorder averages $\overline{\Lambda(x)}=0$,
$\overline{\Lambda(x)\Lambda^\ast(x^\prime)}=aD\delta(x-x^\prime)$
and
\begin{eqnarray}\label{Lam_corr}
\overline{e^{-\Lambda(x)} e^{-\Lambda(x^\prime)}} &=& \int {\cal{D}}\Lambda
\, P[\Lambda] e^{\Lambda(x)} e^{\Lambda(x^\prime)} \nonumber\\
&=& e^{D} + a\delta(x-x^\prime) \left[ e^{2D} - e^{D} \right]\; .
\end{eqnarray}
The last result is derived from the discrete version of the above
functional integral, which leads to $\delta_{ij}e^{2D} +
\left(1-\delta_{ij}\right) e^{D}$ [here
$\delta_{ij}=a\delta(x_i-x_j)$]. Note that the parameter $D$
characterizes the degree of granularity in the SC wire, with
$\sqrt{D}$ proportional to the typical amplitude of spatial
fluctuations in the wire cross section.

The inhomogeneous phase--slip term Eq. (\ref{Hdis}) modifies the
expression for the force operator
\begin{eqnarray}\label{Fe-dis}
F_{ps}^{e} = 8n \: \frac{e} {a^2}\: \frac{v^2} {K} \int dx \: y(x)
\: \sin\left(2\theta \right) \; .
\end{eqnarray}
Substituting in Eq. (\ref{MMee}) and performing the disorder
averaging using the correlation function (\ref{Lam_corr}) we find
\begin{eqnarray}
M_{ee} &\approx & M_{ee}^h+M_{ee}^g \quad\quad {\rm where} \\
M_{ee}^h &= & \frac {4L(4ev^2y_0)^2}{a^4K^2}
e^{D}\int_{-\infty}^{\infty} dx
\int_{0}^{\infty} dt \,t\,{\rm Im}\{C_{ps,1}^{ee}(x,t)\}\; ,\nonumber \\
M_{ee}^g &= & \frac {4L(4ev^2y_0)^2}{a^3K^2} \left[ e^{2D} - e^{D} \right]
\int_{0}^{\infty} dt \,t\,{\rm Im}\{C_{ps,1}^{ee}(0,t)\} \nonumber
\end{eqnarray}
and $C_{ps,1}^{ee}(x,t)$ is given by Eq. (\ref{Cee}) for $n=1$.
This yields the leading $T$--dependence of the resistivity for arbitrary granularity
$D$ in the SC wire:
\begin{eqnarray}\label{rho-final}
\rho(T)&\approx &\frac{4\pi y_0^2}{(2e)^2a}\left[\frac{2\pi aT}{v}\right]^{2/K-3} \\
&\times & \left[A_{h}e^{D}+A_{g}\left( e^{2D} - e^{D} \right)
\left(\frac{2\pi aT}{v}\right)\right]\; , \nonumber \\
A_{h}&=&\frac{\pi^2\Gamma^4(1/2K)}{\Gamma^2(1/K)}\; ,\quad
A_{g}=\frac{\pi^2\Gamma^2(1/K)}{4\Gamma(2/K)} \nonumber \, .
\end{eqnarray}
For extremely weak granularity where $D\ll 1$, the first term in
(\ref{rho-final}) dominates, and the homogeneous result Eq. (\ref{rho-hom-final})
is recovered. However, for $D$ of order 1 or more, the second term is
{\it exponentially enhanced} by the factor $e^{2D}$, yielding
\begin{eqnarray}\label{rho-gran}
\rho(T)&\approx &\frac{4\pi}{(2e)^2a}
A_{g} y_0^2e^{2D} \left[\frac{2\pi aT}{v}\right]^{2/K-2} \, .
\end{eqnarray}
This approximation is consistent with earlier predictions for
granular SC wires\cite{Renn,KP}. Indeed, it indicates that the
phase--slips dominating the resistivity occur at narrow
constrictions in the inhomogeneous wire. Our more general expression
Eq. (\ref{rho-final}) implies that for a fixed $D$, the resistivity
$\rho(T)$ vs. $T$ exhibits a crossover from a ``homogeneous''
power--law behavior [Eq. (\ref{powerlaw}) with $\alpha=2/K-3$] to a
``granular'' limit with a higher exponent $\alpha=2/K-2$. In most
realistic systems we expect this crossover to occur at very low
temperatures: this is in view of the exponential dependence on the
granularity parameter $D$. The typical core action $S_0$ has been
estimated to be of order $10$ or
larger\cite{zaikin,zaikin_prb,DuanCom}; this implies that even small
irregularities ($\sim 10\%$) in the wire diameter corresponding to
$D>1$ are sufficient to enhance the prefactor of the second term in
(\ref{rho-final}) by at least an order of magnitude.

\section{Comparison with Experimental Data}
\label{sec:exp}

As shown in the previous section, the resistivity of a SC wire with
a moderately non--uniform diameter [Eq. (\ref{rho-gran})] far enough
below $T_c$ is expected to be well approximated by a power--law
$T$--dependence of the form (\ref{powerlaw}). In particular, the
exponential enhancement by the granularity parameter $D$ partially
compensates for the exponential suppression by the fugacity
$y=e^{-S_{core}}$, and consequently the prefactor [$\rho_0$ in Eq.
(\ref{powerlaw})] becomes comparable to the observed resistivity in
a number of experiments. In this section we test the relevance to
available experimental data by a direct attempt to fit a power--law.
In the cases where the fit appears to be reasonably good, we can
extract the Luttinger parameter $K$ from the exponent $\alpha=2/K-2$
and compare it to an independent estimate based on the sample
parameters. According to Eq. (\ref{K-def}), $K$ depends in
particular on the wire diameter $d\sim \sqrt{s}$, the capacitance
per unit length $C$ and the superfluid density $n_s$, which can be
expressed in terms of the London penetration depth $\lambda_L$:
\begin{eqnarray}\label{K-dlambda}
K&\approx &\frac{1}{50}\frac{\lambda_L}{d}\frac{A}{\sqrt{C}}  \, .
\end{eqnarray}
Here $A$ is a numerical factor which depends on the geometry of the
wire cross section (e.g., $A=\sqrt{\pi}$ in
Ref.~[\onlinecite{Giordano}] and $A=2$ in Ref.~[\onlinecite{tian}]),
and $C$ is typically of order unity\cite{zaikin} and depends very
weakly (i.e. logarithmically) on $d$; we hereon regard it as a
fitting parameter. The dependence on $\lambda_L$ implies that a
smooth tuning of $K$ is possible by application of a magnetic field
parallel to the wire. Indeed, this was done by Altomare {\it et al.}
(Ref.~[\onlinecite{altomare}]), as will be discussed in more detail
below.

We have considered data obtained in several different experimental
setups: Indium strips studied by Giordano
(Ref.~[\onlinecite{Giordano}]), MoGe wires deposited on carbon
nanotubes studied by the Harvard group
(Ref.~[\onlinecite{tinkham1,tinkham2}]), more recent studies of tin
(Sn) nanowires (Tian {\it et al.}, Ref.~[\onlinecite{tian}]) and
long Aluminum wires (Altomare {\it et al.},
Ref.~[\onlinecite{altomare}]). All of these indicate substantial
deviation from the thermal activation theory \cite{LAMH}, attributed
to QPS. In addition, we note that in most of the SC samples involved
in those experiments the normal resistivity is too low to be
considered in the strictly granular limit (where distinct grains are
weakly coupled), although the wire diameter is likely to be
non--uniform with varying degree of nonuniformity. The samples of
Ref.~[\onlinecite{tinkham1,tinkham2}] exhibit a rather complicated
behavior, following from a number of reasons. First, most of the
wires fabricated in this particular methods are relatively short,
hence finite size effects interfere with the $T$--dependence, and
the dynamics of phase--slips is crucially affected by their
backscattering from the boundaries \cite{oreg}. In addition, some of
the nanotube substrates are not insulating, and their (unknown)
resistance complicates the fitting by additional parameters. Hence,
although some of these samples can be fitted reasonably well by Eq.
(\ref{powerlaw}), a more detailed quantitative analysis has focused
on the other experimental papers.

\begin{figure}[htb]
\begin{center}
\includegraphics[width=8cm,angle=0]{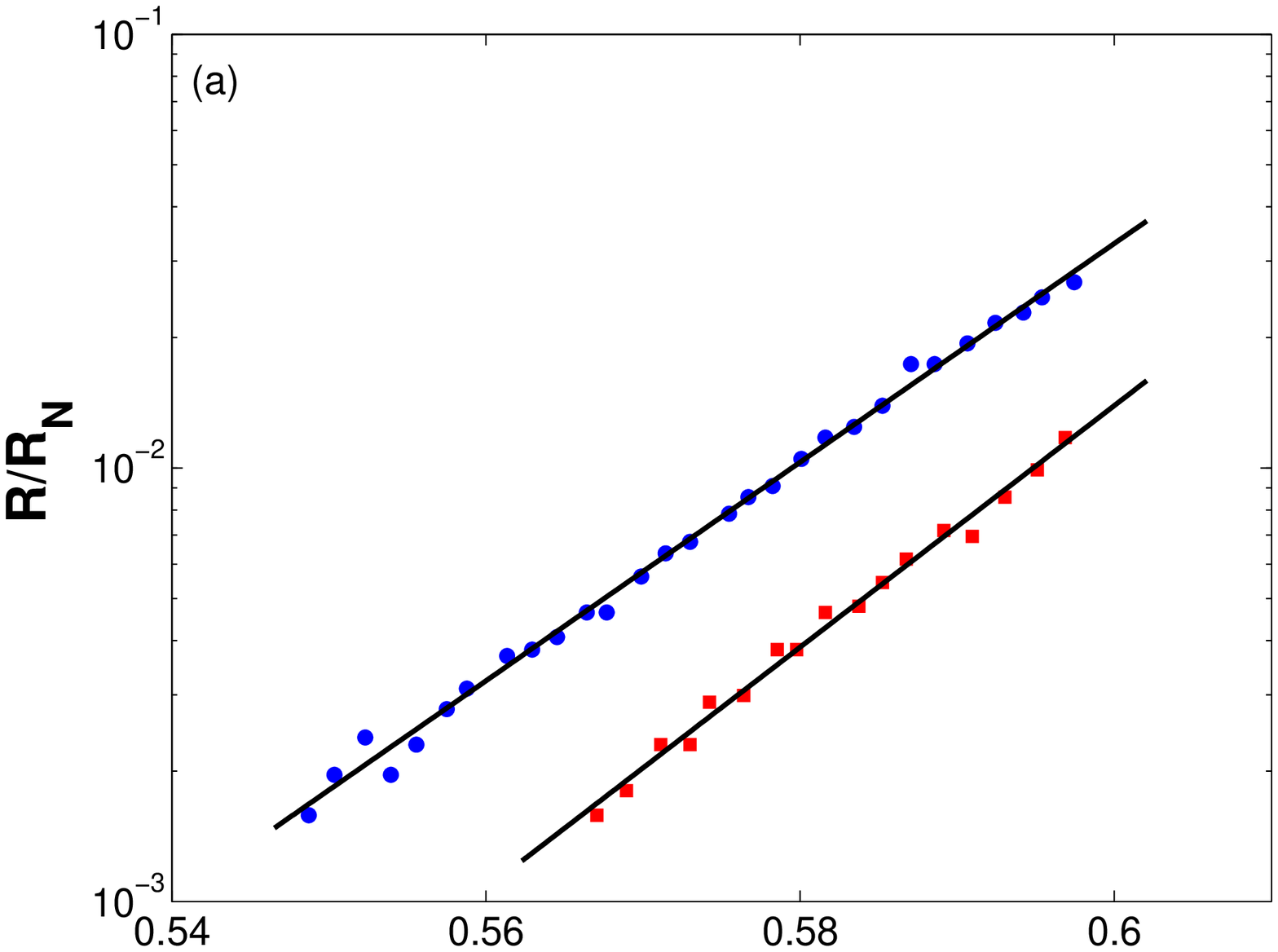}
\includegraphics[width=8cm,angle=0]{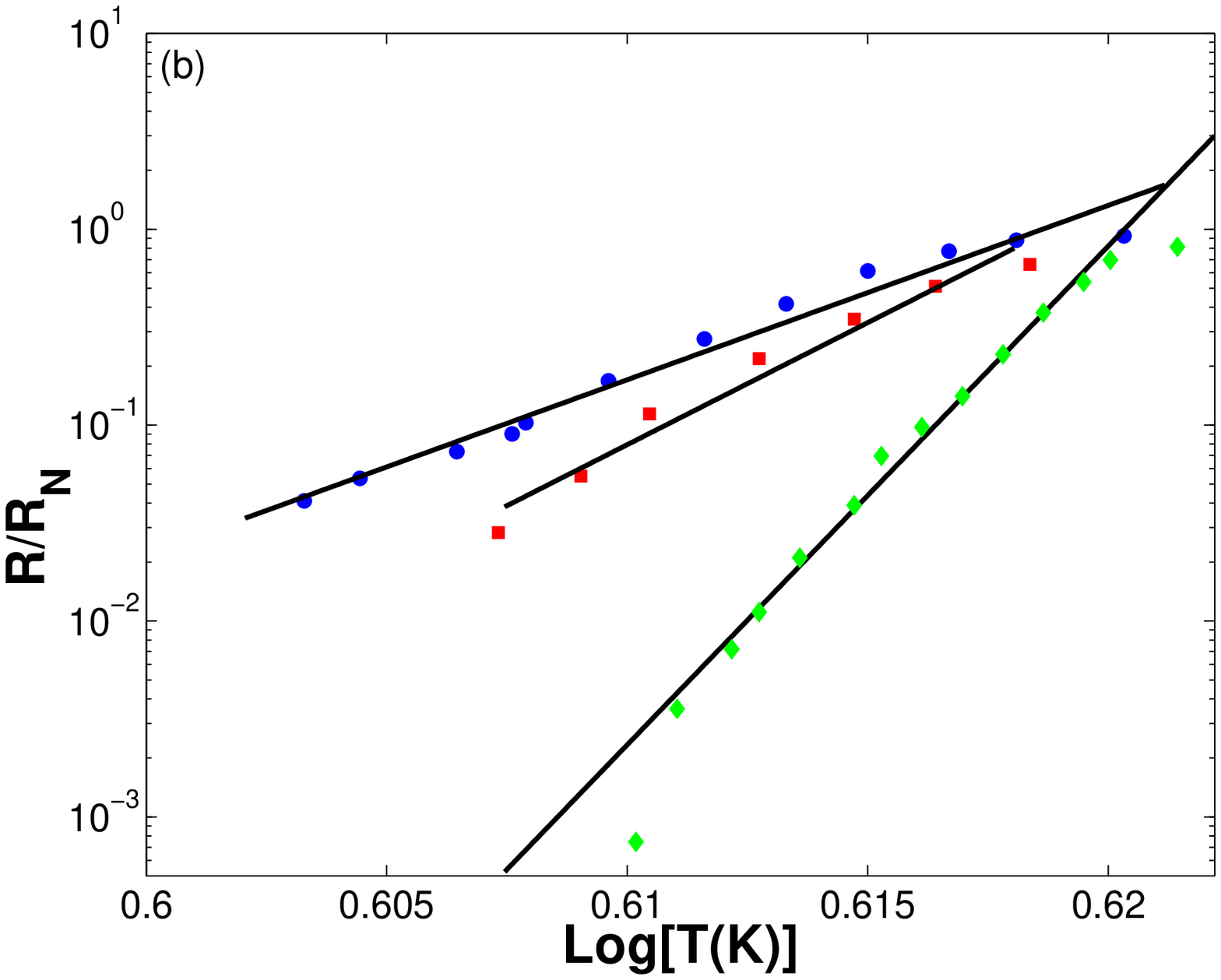}
\caption{Data points obtained from Fig. 1 of
Ref.~[\onlinecite{Giordano}] presenting the resistance vs. $T$,
re--plotted on a log--log scale; the solid lines correspond to
fitting functions of the form $T^\alpha$. (a) In the low--$T$
(QPS) regime: blue circles (red squares) correspond to the samples
of diameter $410\AA$ ($505\AA$); the corresponding exponents are
$\alpha=25.2\pm 0.2$ and $\alpha=27.85\pm 0.4$, respectively. (b)
The $410\AA$ and $505\AA$ diameter samples in the high--$T$
regime, and all data points of the $720\AA$ diameter sample (green
diamonds); $\alpha=89\pm 13$, $\alpha=125\pm 8$ and $\alpha=260\pm
7$, respectively.} \label{fig:gio}
\end{center}
\end{figure}
We first consider the data of Ref.~[\onlinecite{Giordano}]. Fig.
(\ref{fig:gio}) presents the resistance of two samples as a
function of $T$ on a log--log scale. Note that the length of the
shortest wire studied is 80 $\mu$m. In comparison, the effective
length set by the temperature scale $L_T = (v/T)$ for the relevant
$T\sim 4\,K$ is of order 10 $\mu$m; this follows from the estimate
$v\sim c d/\lambda_L$, where $c$ is the velocity of light and the
measured $\lambda_L\approx 1300\AA$ obtained for Indium films of
comparable thickness \cite{toxen}. The wires therefore fulfill the
long wire condition $L\gg L_T$, and can be considered good
candidates for testing the scaling of $\rho(T)$ with $T$. In the
low $T$ regime [Fig. (\ref{fig:gio}(a))], the fit to a power--law
is very good\cite{realTc}. In addition, the values of $\alpha$
corresponding to the two samples of diameters $410\AA$ and
$505\AA$ yield the Luttinger parameters $K = 0.0735$ and $K =
0.067$, respectively. These are consistent with the values
obtained by inserting the relevant $d$ and $\lambda_L$ in Eq.
(\ref{K-dlambda}), provided $C\approx 2$.

In principle, one may argue that a fit by a power--law with a high
exponent (especially within a limited range of $T$) is not easily
distinguishable from an exponential function. However, to put the
above analysis to a further test, we tried to fit the data
obtained in the high $T$ regime (in the close vicinity of $T_c$)
to a power--law as well. In this regime, there is no justification
for such a functional dependence on $T$, as the thermal activation
theory for phase--slips\cite{LAMH} is expected to work much
better. Indeed, the result depicted in Fig. (\ref{fig:gio}(b))
indicates an obvious failure of a $T^\alpha$ trial function: the
best fit to a straight line on a log--log plot yields an
anomalously large exponent ($\alpha$ ranging from $89$ to $260$),
and a systematic deviation from the fitting function. This is in
sharp contrast with the situation in the low--$T$ regime, and
therefore strengthen our confidence that in the latter case, the
analysis is valid and provides a meaningful confirmation of the
theoretical prediction.
\begin{figure}[htb]
\begin{center}
\includegraphics[width=8cm,angle=0]{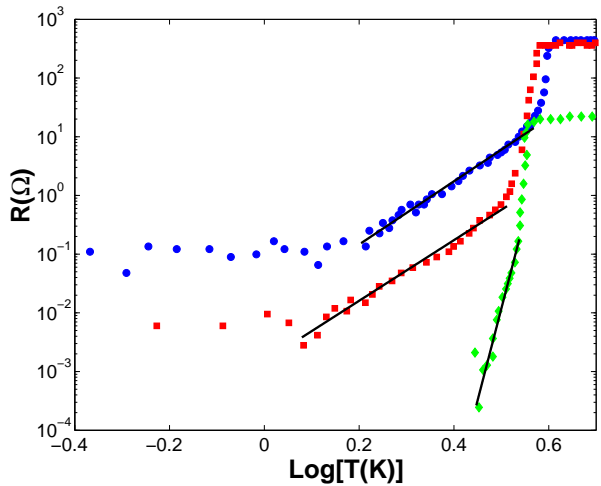}
\caption{Data points obtained from Fig. 1 of
Ref.~[\onlinecite{tian}] presenting the resistance vs. $T$,
re--plotted on a log--log scale: sample diameters are 20 nm (blue
circles), 40 nm (red squares), and 60 nm (green diamonds). The
solid lines correspond to fitting functions of the form
$T^\alpha$, with $\alpha=5.38\pm 0.2$, $\alpha=5.16\pm 0.2$ and
$\alpha=31.2\pm 1.3$, respectively.} \label{fig:tian}
\end{center}
\end{figure}

We next consider the data of Tian {\it et al}
(Ref.~[\onlinecite{tian}]). The results obtained from resistivity
measurements in three different tin (Sn) nanowires of diameters
ranging between 20 nm and 60 nm are depicted in Fig.
(\ref{fig:tian}). Similarly to Giordano's data, in all three samples
a ``kink" is observed at some temperature $T^\ast<T_c$, below which
the decrease of $R$ as $T$ is lowered becomes more moderate. In the
two thinner samples, the low--$T$ section of the data is quite
noisy, and $R(T)$ seems to saturate. This behavior is possibly due
to serial contribution from the contacts. Otherwise, however, the
power--law dependence appears to be a good fit for $T<T^\ast$.

To test the validity of Eq. (\ref{K-dlambda}) in this system, one
requires an independent estimate of $\lambda_L$. A direct
measurement is not available, however, an indirect estimate of the
ratio $\lambda_L/d$ based on a measurement of the effective critical
magnetic field\cite{Liu} yields $\lambda_L/d\approx 1.5$ for
$d\approx 70$ nm, which is close in diameter to the thickest sample
of Fig. (\ref{fig:tian}). This is consistent with Eq.
(\ref{K-dlambda}) for $C\approx 0.98$, a quite reasonable estimate
for the capacitance. Unfortunately, reliable data on $\lambda_L/d$
in the thinner samples is not available. A naive extrapolation of
the $d$--dependence of $\lambda_L$ to lower values of $d$ based on
the Landau--Ginzburg theory\cite{SCbook} ($\lambda_L\sim
1/\sqrt{d}$) yields $K=0.315$ for the $20$ nm sample, a reasonable
approximation to the value of $K$ obtained from the fit ($K=0.274$),
however the expected scaling with $d$ fails in the case of the $40$
nm sample. Indeed, as shown explicitly in Ref.~[\onlinecite{Liu}],
the naive scaling $\lambda_L(d)$ works well above $60$ nm, but
breaks down for the $d=40$ nm wire (no data is given for even
thinner samples).

We finally focus our attention on the most recent experimental work
of Altomare {\it et al.} (Ref.~[\onlinecite{altomare}]). This group
has studied long and thin Aluminum wires, which appear to be ideal
candidates for comparison with the QPS theory. In addition, the
application of a magnetic field $H$ enables a continuous tuning of
the superfluid density in a single sample of fixed $d$. Following
Eq. (\ref{K-def}), we therefore expect the $H$--dependence of $K$ to
be related to $n_s$ via $1/K(H) \sim \sqrt{n_s(H)}$. The functional
dependence of $n_s(H)$ on $H$ can be derived from a simple
calculation using the Landau--Ginzburg theory at finite magnetic
field\cite{SCbook}. This yields
\begin{eqnarray}\label{ns-H}
n_s(H)=n_s(0)\left(1-\frac{H^2}{H_{\parallel}^2}\right)\; ,\quad
H_{\parallel}\equiv\frac{\mathcal{N}\lambda_L}{d}H_c
\end{eqnarray}
where $H_c$ is the bulk critical field, and $\mathcal{N}$ a
numerical factor which depends on the geometry: e.g.,
$\mathcal{N}=2\sqrt{6}$ in a thin film of thickness $d$, and
$\mathcal{N}=8$ in a cylindrical wire of diameter $d$.
$H_{\parallel}$ marks an estimated critical field where the inverse
Luttinger parameter is expected to vanish according to
\begin{eqnarray}\label{KvsH}
\frac{1}{K(H)}=\frac{1}{K(0)}
\left(1-\frac{H^2}{H_{\parallel}^2}\right)^{1/2}\; .
\end{eqnarray}
\begin{figure}[htb]
\begin{center}
\includegraphics[width=8cm,angle=0]{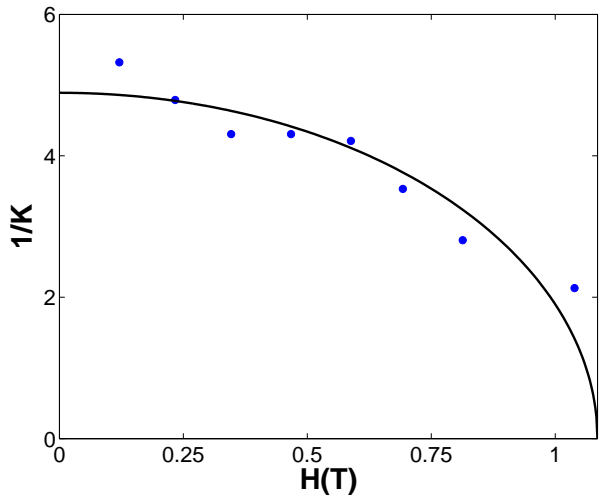}
\caption{Data obtained from the unpublished notes
(Ref.~[\onlinecite{altomare}]), plotted in terms of the inverse
Luttinger parameter ($1/K$) as a function of magnetic field $H$. All
data points correspond to the lowest temperature measurements
($T=0.4$ K). The solid curve corresponds to the formula Eq.
(\ref{KvsH}), with $H_{\parallel}\approx 1.1\,{\rm T}$.}
\label{fig:altomare}
\end{center}
\end{figure}

The unpublished notes included in Ref.~[\onlinecite{altomare}]
enable a direct comparison of the experimental data to the above
prediction. The experimental results include the Ohmic resistance as
well as current--voltage characteristics which include a non--linear
part. In the notes, it is shown that the latter can be fitted to a
power--law $V\sim I^\nu$. This is suggested by the authors as a
plausible alternative to the phenomenological exponential expression
used as a fitting function in the published Letter. Such power--law
in the $I-V$ characteristic is consistent with a $T$--dependence of
the Ohmic resistance of the form $R(T)\sim T^\alpha$, with
$\alpha=\nu -1$. The exponent $\nu$ is then extracted from the data
for different values of the magnetic field, and plotted as a
function of $H$. Using the linear relation of $\nu$ to $1/K$, we
plot the data in the form depicted in Fig. (\ref{fig:altomare}),
where it is fitted to a trial function of the form (\ref{KvsH}). The
fit is reasonably successful, and is better than the $H$--dependence
derived from Giordano's formula\cite{altomare}. It yields a critical
field $H_{\parallel}\approx 1.1\,{\rm T}$.

To compare with an independent estimate of $H_{\parallel}$ based on
the experimental parameters, we use the dirty limit expression for
the bulk critical field\cite{SCbook} $H_c=\phi_0/\pi
2^{2/3}\lambda_L\xi$ (with $\xi$ the SC coherence length in the
dirty limit). Employing Eq. (\ref{ns-H}) we observe that, up to the
numerical constant $\mathcal{N}$, $H_{\parallel}$ is essentially
determined by $d$ and $\xi$ only. The experimental parameters
mentioned in Ref.~[\onlinecite{altomare}] imply $d\approx 60\AA$,
$\xi\approx 1280\AA$ (the latter, however, relies on the text--book
expression for $\xi$ in a dirty Aluminum and the measured mean free
path). The wires are actually thin strips with a rectangular cross
section, hence $\mathcal{N}$ cannot be determined accurately, but is
expected to be intermediate between a thin film and a wire, yielding
$H_{\parallel}$ between $1.4\,{\rm T}$ and $2\,{\rm T}$. The order
of magnitude is consistent with the value extracted from Fig.
(\ref{fig:altomare}). Considering the fact that the above estimate
of $H_{\parallel}$ does not rely on any fitting parameters, this is
a reasonably good agreement.

\section{Conclusions}
\label{sec:conclude}

We have studied the low $T$ resistivity of a dirty SC wire using a memory formalism
approach. This method allows a perturbative treatment of corrections to the low energy
effective theory, describing the dynamics of phase fluctuations in the SC order parameter.
In the limits of either ideally homogeneous wire or strongly inhomogeneous (granular) wire,
our results for $\rho(T)$ recover the power--law behavior $\sim T^\alpha$ derived in
earlier theoretical literature using the instanton technique. We show that more generally,
the expression for $\rho(T)$ interpolates between the two limits, where the relative
weights of each are smoothly tuned by a ``granularity parameter'' $D$ describing the
typical fluctuations in the diameter of the wire. In particular, the inhomogeneity of the
wire leads to an exponential enhancement of the phase--slips induced resistivity.

Following these calculations, we infer that the low energy theory is
likely to be relevant to experimental measurements of $\rho(T)$ in
the low $T$, QPS regime. We directly test the validity of the
power--law ansatz as a fit to experimental data obtained in a
variety of different samples, and found a very good agreement in
sufficiently long wires. The exponent $\alpha$ extracted from the
$\rho(T)$ data is found to be consistent with the values obtained
from an independent estimate, based on known parameters of the
corresponding samples (as much as the information was available). In
addition, its dependence on a parallel magnetic field is consistent
with the theoretical expectation. We conclude that the low--$T$
theory for dirty SC wires provides a plausible interpretation of
data presented in the experimental literature, which is better
justified than previously used phenomenological effective models.
Our conclusions will hopefully encourage a further, more systematic
investigation of the comparison between theory and experiment.

\acknowledgements

We wish to thank Konstantin Arutyunov, Sergei Khlebnikov, Ying Liu,
Yuval Oreg, Gil Rephael, Dan Shahar, Nayana Shah and especially
Achim Rosch for numerous useful discussions. In addition, we thank
Konstantin Arutyunov, Ying Liu and M. Tian for showing us
unpublished data. This work was partially supported by a grant from
GIF, the German Israeli Foundation for Scientific Research and
Development.

\appendix

\section{Derivation of the Resistivity within the Memory Function Approach}
\label{sec:memory}

In this Appendix we review the general aspects of the memory matrix
approach\cite{forster,woelfle,giamarchi,rosch}, which turns out to
serve as a useful tool for evaluating transport properties in
systems where approximate conservation laws can be easily
identified. As we show below, the calculation of d.c. electric
resistivity of a SC wire provides a particularly simple example for
the application of this approach, which amounts to a straightforward
perturbative treatment of any processes responsible for relaxing the
transport current.

At finite but low $T$, the fixed point Hamiltonian $H_0$ [Eq.
(\ref{H0})] provides the leading contribution to thermodynamic
properties of the SC wire. However, by itself it does not give
access to transport properties: being a translationally invariant
integrable model, it possesses an infinite number of conserved
currents, including in particular the electric charge transport
current. Since the current cannot degrade, the d.c. conductivity is
infinite even for $T>0$. To get the leading non--trivial
contribution to transport, it is therefore necessary to add the
irrelevant corrections, leading to a slow but finite relaxation rate
of the currents. In our case, the prominent corrections are the
phase--slips terms $H_n^{ps}$ [Eq. (\ref{Hps})]: note that other
irrelevant terms, e.g., of the form
$\int(\partial\phi)^m(\partial\theta)^n$, $\int(\partial^m \phi)
(\partial\phi)^n$ with $n+m>2$ as well as pair breaking terms, which
are already neglected in Eq. (\ref{HSG}), do not contribute
significantly to transport\cite{rosch,SAR}. We then evaluate the
resistivity $\rho(T)$ employing the memory matrix approach, which
takes advantage of the fact that the memory matrix -- a matrix of
decay--rates of the slowest modes in the system -- is perturbative
in the irrelevant operators. This is in contrast with the
conductivity matrix, which is a highly singular function of these
perturbations.

A crucial step in the derivation of transport coefficients in the
memory matrix approach is the identification of primary slowly
decaying currents of the system. These are conserved ``charges'' of
the fixed point Hamiltonian $H_0$, whose conservation is slightly
violated by certain irrelevant perturbations. In the case of the
Luttinger model \cite{rosch}, these include the charge current $J_e$
[Eq. (\ref{Je:main})] and, in addition, the total translation
operator
\begin{equation}\label{JT}
J_T=-\int dx \Pi\partial_x \phi\; .
\end{equation}
In the above, we approximate the currents by contributions from the
collective degrees of freedom (phase fluctuations) only.
Non--superconducting contributions associated with unpaired
electrons are exponentially suppressed as $e^{-\Delta/T}$ (with
$\Delta\sim T_c$) for $T\ll T_c$. The correlators of $J_e$, $J_T$
determine the conductivity matrix $\hat\sigma(\w,T)$ at frequency
$\w$ and temperature $T$ via the Kubo formula:
\begin{eqnarray}\label{sigma-def}
\sigma_{pq}(\w,T) =\frac{1}{TL}\int_0^{\infty} dt e^{i \omega t}
\left(J_p(t)|J_q\right) \; ,
\end{eqnarray}
where following Ref.~[\onlinecite{forster}] we have introduced the
scalar product (of any two operators $A$ and $B$)
\begin{eqnarray}
\left(A(t)|B\right)&\equiv& T \int_0^{1/T} d\lambda \left\langle
A(t)^\dagger B(i \lambda) \right\rangle\; .
\end{eqnarray}
The d.c. electrical resistivity is then given by
$\rho(T)=1/\sigma_{ee}(T)$, where $\hat{\sigma}(T)$ is the
$\w\rightarrow 0$ limit of $\sigma_{ee}(\w,T)$. However, a direct
application of Eq. (\ref{sigma-def}) at low $T,\w$ is rather subtle:
since $[J_q,H_0]=0$ (for $q=e,T$), the relaxation rate of the
currents $\partial_t J_q=i[J_q,H]$ is dictated by the irrelevant
corrections, hence tends to vanish in the limit $T\rightarrow 0$.
This leads to divergences in the conductivities, since the currents
do not decay in this limit.

To enable a controlled  perturbative expansion in the relaxation
rates $\partial_t J_q$, we therefore recast the conductivity matrix
in terms of  a memory matrix $\hat{M}$:
\begin{eqnarray} \label{sigma-t}
\hat\sigma(\w,T)&=&\hat{\chi}(T) \left(\hat{M}(\w,T)- i \w
\hat{\chi}(T) \right)^{-1}\hat{\chi}(T),
\end{eqnarray}
in which
\begin{eqnarray}\label{M}
{M}_{pq}(\w,T) \equiv \frac{1}{TL} \left(\partial_t J_p \left|
{\mathcal Q} \frac{i}{\w-{\mathcal Q}{\mathcal L}{\mathcal Q}}
{\mathcal Q} \right| \partial_t J_q \right)
\end{eqnarray}
and $\hat{\chi}$ is the matrix of static susceptibilities
\begin{eqnarray}\label{chi-def}
{\chi}_{pq}= \frac{1}{T L} (J_p|J_q)\; .
\end{eqnarray}
Here ${\mathcal L}$ is the Liouville operator defined by ${\mathcal
L}A = [H,A]$, and ${\mathcal Q}$ is the projection operator on the
space perpendicular to the slowly varying variables $J_q$,
\begin{eqnarray}\label{Q}
{\mathcal Q}=1-\sum_{pq} |J_p) \frac{1}{T L} (\hat{\chi}^{-1})_{pq}
(J_q|\; .
\end{eqnarray}
Note that similarly to Ref.~[\onlinecite{SAR}], we choose a
convenient definition of the memory matrix [Eq. (\ref{M})], which is
slightly different than the standard literature. The perturbative
nature of $\hat{M}$ is transparently reflected by this expression:
in particular, the operators $\partial_t J_q$ are already linear in
the irrelevant corrections to $H_0$. This enables a systematic
perturbative expansion of the correlator in Eq. (\ref{M}) in the
small parameter characterizing the relative size of the corrections
-- in our case, the exponential factors $y^n$ [see Eq. (\ref{Hps})].

We now obtain an expression for the d.c. electric resistivity by
setting $\w=0$ in Eq. (\ref{sigma-t}). This yields
\begin{eqnarray}\label{rho-M}
\rho(T)= \frac{M_{ee}M_{TT}-M_{eT}^2}
{\chi_{ee}^2M_{TT}+2\chi_{eT}\chi_{ee}M_{eT}+\chi_{eT}^2M_{ee}}\, ,
\end{eqnarray}
where $\hat{M}$ is evaluated from Eq. (\ref{M}) in the limit
$\w\rightarrow 0$. (Note that here we have used the fact that both
matrices $\hat{M}(T)$ and $\hat{\chi}(T)$ are symmetric.) The
relaxation rate operators $\partial_t J_q$ are given by
\begin{eqnarray}\label{dtJ}
\partial_t J_q=\sum_n F_{ps,n}^q\; ,
\end{eqnarray}
where we have defined the `force' operators
\begin{eqnarray}\label{Fps-def}
F_{ps,n}^q\equiv i[J_q,H_n^{ps}]\; .
\end{eqnarray}
To leading order in the perturbations $H_n^{ps}$, Eq. (\ref{M}) for
$\hat{M}$ is greatly simplified: one can set ${\mathcal L}={\mathcal
L}_{0}$ with ${\mathcal L}_0=[H_0,.]$ and ${\mathcal Q}=1$. This
yields
\begin{eqnarray}\label{MM}
\hat{M}(T) \approx \sum_{n} \hat{M}_{ps,n}\; ,
\end{eqnarray}
where the elements of $\hat{M}_{ps,n}$ are given by
\begin{eqnarray}\label{MMM}
 M_{ps,n}^{pq}\equiv \lim_{\w\rightarrow 0}\frac{
\langle F_{ps,n}^p;F_{ps,n}^q \rangle^0_\w- \langle
F_{ps,n}^p;F_{ps,n}^q \rangle^0_{\w=0}}{i \w}
\end{eqnarray}
in which $\langle F^p;F^q\rangle^0_{\w}$ is the retarded correlation
function evaluated with respect to $H_{0}$.

Substituting Eqs. (\ref{JT}) and (\ref{Hps}) in Eq. (\ref{Fps-def})
and taking the $L\rightarrow \infty$ limit, it is easy to see that
$F_{ps,n}^T=0$. Indeed, this follows from the translational
invariance of $H_n^{ps}$. As a result, $M_{TT}$ and $M_{eT}$
identically vanish. This implies that {\em if} both $\chi_{eT}$ and
$M_{ee}$ do not vanish, Eq. (\ref{rho-M}) yields $\rho(T)=0$ {\em
even at finite $T$}; namely, the creation of free phase--slips
appears to be insufficient to generate a finite dissipation. Such a
result cannot be reconciled with our understanding, that dissipation
should occur at the normal cores during a phase--slips event, where
the wire behaves temporarily as a normal dirty metal. Indeed, we
should recall that the translational invariance of $H_n^{ps}$ is not
of microscopic origin; rather, these terms in the effective
Hamiltonian are obtained after averaging over disorder in the
electron system\cite{zaikin_prb}. Their form reflects the total
absorption of the finite momentum $P=2\pi sn_s$ generated in a
phase--slip process by the core electrons. In a clean SC wire, such
momentum transfer is not effective, leading to an exponential
suppression of $M_{TT}$, $M_{ee}$ and yielding a finite but
exponentially small resistivity $\rho(T)\sim e^{-vP/2T}$ (see Ref.
~[\onlinecite{khleb1,KP}]). Stronger disorder in the underlying
electronic system should only enhance the resistivity, and not vice
versa! This apparent paradox is resolved by the remarkable
observation\cite{RA_JLTP} that, following the commensurate relation
between the total momentum $P$ and density per unit length,
$\chi_{eT}$ actually {\em vanishes identically}, even at finite $T$.
Setting $\chi_{eT}=0$ in Eq. (\ref{rho-M}), we then find that the
expression for the resistivity $\rho(T)$ reduces to the simplified
form Eq. (\ref{rho-ee}).

The above arguments are accurate as long as indeed the currents
$J_e$, $J_T$ are given by Eqs. (\ref{Je:main}),~(\ref{JT}), i.e.,
when normal electron contributions to the currents are neglected.
However, if such normal contributions are not negligible, they would
at the same time modify the translation invariance of the
Hamiltonian, and all entries in Eq. (\ref{rho-M}) would be finite.
Similarly, translational invariance is explicitly broken once we
account for random spatial fluctuations in the SC wire cross
section, by introducing the inhomogeneous phase--slip term Eq.
(\ref{Hdis}). This modifies the force operator $F_{ps}^{e}$ [now
given by Eq. (\ref{Fe-dis})], and induces a non--trivial
contribution to $F_{ps}^{T}$:
\begin{eqnarray}\label{FeT}
F_{ps}^{T}= -i \frac{2} {\pi}\: v^2 \int dx\: y(x)\:
\left(\partial_x \theta\right)\: \sin \left(2 \theta
\right)\nonumber\; .
\end{eqnarray}
However, the resulting matrix $\hat{M}(T)$ obtained after
substitution in the approximate form Eq. (\ref{MMM}) is still
diagonal: to see this, we note that the off-diagonal element is
given (after disorder averaging) by
\begin{eqnarray}\label{MeT}
& M_{eT}&\approx \frac{4evLy_0^2}{ K} \\
&\times &\int_{0}^{\infty} dt \, t\, \int_{-\infty}^{\infty} dx\;
\overline{e^{-\Lambda(x)} e^{-\Lambda(0)}}\; {\rm
Im}\{C_{eT}(x,t)\}\nonumber
\end{eqnarray}
where
\begin{eqnarray}\label{CeT}
C_{eT}(x,t)&=&2\partial_xG_\theta(x,t)e^{4G_\theta(x,t)}\; ,
\end{eqnarray}
in which $G_\theta(x,t)$ is given by Eq. (\ref{Green-T0}). It is
apparent from Eqs. (\ref{CeT}) and (\ref{Green-T0}) that
$C_{eT}(x,t)$ is an antisymmetric function of $x$, and hence the
integral over $x$ in Eq. (\ref{MeT}) vanishes, yielding $M_{eT}=0$.
The static susceptibility matrix [Eq. (\ref{chi-def})] evaluated to
leading order in the perturbation $H_{sp}$ is also approximately
diagonal, with $\chi_{ee}$ given by Eq. (\ref{chi}). We therefore
find that the leading contribution to Eq. (\ref{rho-M}) for the
resistivity again practically reduces to Eq. (\ref{rho-ee}).


\begin{thebibliography}{33}
\bibitem{realTc}
It should be pointed out that the crossover temperature $T^*$ is not
too far from the value of $T_c$ as quoted by Giordano in
Ref.~[\onlinecite{Giordano}], which corresponds to the thin film
critical temperature, assumed to apply for the wire as well. As a
consequence, one might expect a $T$--dependence of the exponent
$\alpha$ associated mainly with the $T$--dependence of $\lambda_L$
near $T_c$. Such deviation from a pure power--law is, however, not
supported by the experimental data for $T<T^*$. A possible
resolution for that might be that the actual $T_c$ in a wire and
film are actually quite different. Indeed, $T_c$ depends crucially
on dimensionality: for example, it is significantly enhanced in a
thin film compared to the bulk (three--dimensional) material. In
practice, $T_c$ is often treated as a fitting parameter (see, e.g.,
Ref.~[\onlinecite{oreg}]).
\end{thebibliography}
\end{document}